\documentclass[10pt]{article}
\usepackage[OE]{express}
\usepackage{mathrsfs}

\begin{document}
\title{Loss-induced transparency in optomechanics}

\author{H. Zhang,\authormark{1} F. Saif,\authormark{2} Y. Jiao,\authormark{1} and H. Jing\authormark{1,*}}

\address{\authormark{1}School of Physics and Electronics, Hunan Normal University, Changsha 410081, China\\
Key Laboratory of Low-Dimensional Quantum Structures and Quantum Control of Ministry of Education, Department of Physics and Synergetic Innovation Center for Quantum Effects and Applications, Hunan Normal University, Changsha 410081, China\\
\authormark{2}Department of Electronics, Quaid-i-Azam University, 45320 Islamabad, Pakistan}

\email{\authormark{*}jinghui73@foxmail.com} 



\begin{abstract}
We study optomechanically induced transparency (OMIT) in a compound system consisting of coupled optical resonators and a mechanical mode, focusing on the unconventional role of loss. We find that optical transparency can emerge at the otherwise strongly absorptive regime in the OMIT spectrum, by using an external nanotip to enhance the optical loss. In particular, loss-induced revival of optical transparency and the associated slow-to-fast light switch can be identified in the vicinity of an exceptional point. These results open up a counterintuitive way to engineer micro-mechanical devices with tunable losses for e.g., coherent optical switch and communications.
\end{abstract}



\section{Introduction}

Cavity optomechanics (COM) \cite{Aspelmeyer2014CavityOptomechanics,Metcalfe2014Applications}, viewed as a new milestone \cite{Milestone2006} in the history of optics, has significantly extended fundamental studies and practical applications of coherent light-matter interactions. A wide range of COM devices, such as  solid-state resonators \cite{Optomechanicalcavity2008Kippenberg}, atomic gases \cite{Atomicgas2008Brennecke}, or liquid droplets \cite{Droplet2016Dahan}, has been created for diverse purposes. Important applications of these devices \cite{Metcalfe2014Applications} include quantum transducer \cite{QtOptoe2018Midolo,Qt2010Rabl,Qt2013Bochmann}, mechanical squeezing \cite{Squeezing2015Wollman}, phonon lasing \cite{PhononLaser2010Grudinin,PhononLaser2014Jing,EPPhononLaser2017Lv}, and ultra-sensitive motion sensing \cite{Sensing2010Krause,Measurement2012Gavartin}. Another intriguing example closely related to the present work is optomechanically induced transparency (OMIT) \cite{OMIT2010Agarwal,RAonOMIT2017LiuYongChun,OMIT2010Weis,OMIT2011Safavi,OMIT2011Teufel,OMIT2013Karuza,CascadedOMIT2015Fan,OMIT2014Dongchunhua,CompensationKerr2016Shen}, as already demonstrated in e.g., a microtoroid resonator \cite{OMIT2010Weis}, a crystal-nanobeam system \cite{OMIT2011Safavi}, a microwave circuit \cite{OMIT2011Teufel}, a membrane-in-the-middle cavity  \cite{OMIT2013Karuza}, a cascaded COM device \cite{CascadedOMIT2015Fan}, an optical cavity coupled to Bogoliubov mechanical modes \cite{OMIT2014Dongchunhua}, and a nonlinear Kerr resonator \cite{CompensationKerr2016Shen}. OMIT is generally viewed as an analog of electromagnetically induced transparency (EIT) well-known in atomic physics \cite{EIT1991Boller,EITScully}, i.e., arising due to destructive interference of two absorption channels of the probe photons (by the cavity or the phonon mode). Beyond this picture, novel effects have also been revealed, such as nonlinear OMIT \cite{NonlinearOMIT2013Marquardt,HigherOrderOMIT2012Xiong,HigherOrderKerr2018Jiao,HigherOrderOMIT2016Jiao}, two-color OMIT \cite{TwoColorOMIT2014Wang,MultiOMIT2018Ullah}, nonreciprocal OMIT \cite{SpiningOMIT2017Lv}, and reversed OMIT \cite{PTOMIT2015Jing}. Promising applications of OMIT devices are actively explored as well, such as optical memory \cite{OMIT2011Safavi,Lightstorage2013Fiore}, phononic engineering \cite{OMITCooling2014Guo,OMITCooling2014Ojanen,OMITCooling2015Liu}, and precision measurements \cite{PrecisionMeasurement2012Zhang,PrecisionMeasurement2017Xiong}.

In this work, we study the unexpected role of loss in OMIT with a compound COM system. We note that in comparison with single-cavity devices, coupled-cavity COM has several unique properties enabling more advantages in applications. For example, both the input light and its frequency sideband can be resonantly tuned to achieve efficient phonon cooling \cite{CoupledCavity2012Fan}. The inter-cavity coupling strength, strongly affecting the circulating power in the resonators, also provides a tunable parameter to realize e.g., phonon lasing \cite{PhononLaser2010Grudinin,PhononLaser2014Jing,EPPhononLaser2017Lv}, unconventional photon blockade \cite{UCPB2013Xu}, reversed OMIT \cite{PTOMIT2015Jing}, and highly-efficient optical control \cite{OMITE2016Li}. In particular, by coupling an active (e.g., Erbium ion-doped) resonator to a lossy one \cite{PTdevice2014Peng,PTdevice2014Chang}, COM devices with an exceptional point (EP) \cite{PhononLaser2014Jing,PTOMIT2015Jing,PTChaos2015LvXin,PTCOM2016Xu,EPPhononLaser2017Lv,HighOrderEPPhononCooling2017Jing}, featuring non-Hermitian coalescence of both eigenvalues and eigenfunctions \cite{PT1998Bender,PT2007Bender,PT2013Bender}, can be created. In view of novel functionalities enabled by EP devices, e.g., single-mode lasing or ultrahigh-sensitive sensing \cite{PTReview2016Konotop,PTReview2018ELGanainy,PTReview2017Feng}, this opens up a new route to operate COM devices at EPs for various applications.

Here we probe the EP features in OMIT, without using any active gain, but increasing the optical loss \cite{LIT2009Guo,LITandEP2014Peng}. We note that in a recent experiment \cite{LITandEP2014Peng}, by placing an external nanotip near a microresonator and thus increasing the optical loss, counterintuitive EP features, i.e., suppression and revival of lasing were demonstrated \cite{LITandEP2014Peng}. Similar EP features in optical transmissions, i.e., loss-induced transparency (LIT) were reported previously in a purely optical experiment \cite{LIT2009Guo}. Our purpose here is to show the LIT features in OMIT devices, i.e., loss-induced suppression and revival of optical transparency at the EP. In addition, we find that by increasing the optical loss, strong absorption regimes in conventional OMIT can become transparent, accompanying by a slow-to-fast light switch in the vicinity of the EP (for similar reversed-OMIT features, see also Ref.\,\cite{PTOMIT2015Jing} in an active COM system). The unconventional role of loss on the higher-order OMIT sidebands \cite{HigherOrderOMIT2012Xiong,HigherOrderOMIT2016Jiao,HigherOrderKerr2018Jiao} is also probed. These results indicate a counterintuitive way to achieve optical switch and communications with OMIT devices, without the need of any active gain or complicated materials.

\section{Model and results}

\begin{figure}[htbp]
\centering
\includegraphics[width=13 cm]{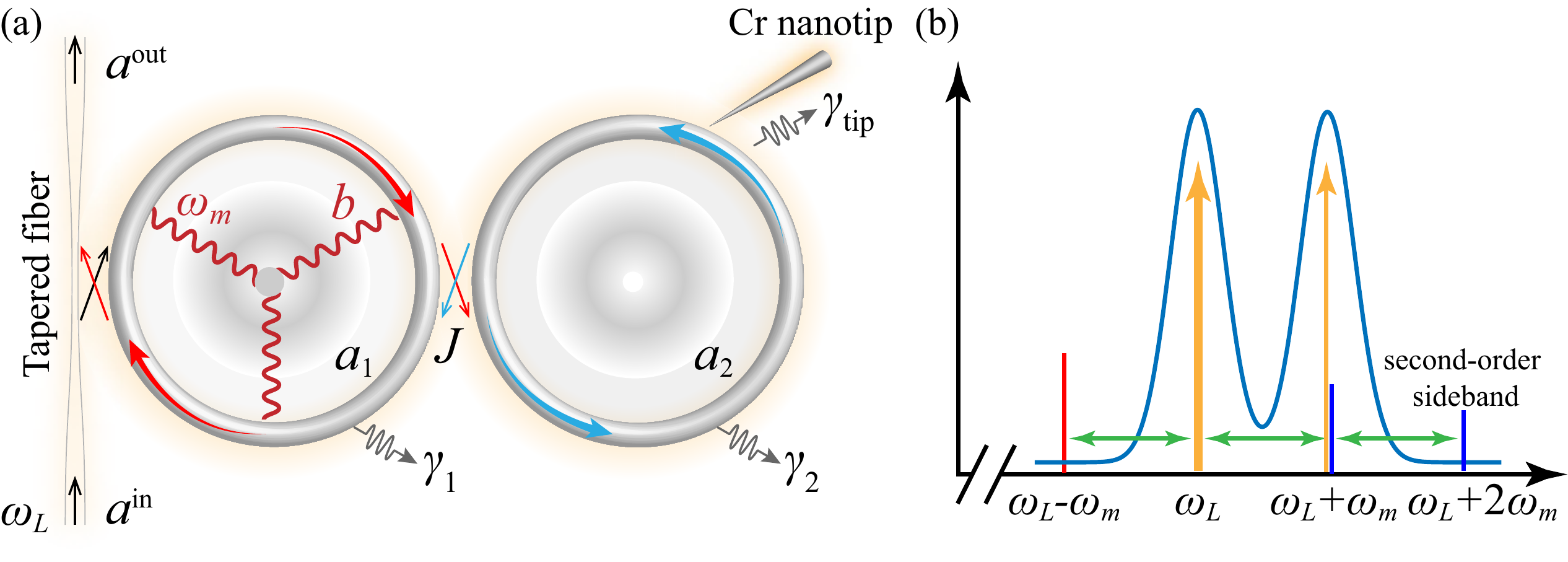}
\caption{(a) Schematic diagram of the compound COM system, with an additional optical loss $\gamma_{\text{tip}}$ induced by a Cr-coated nanofiber tip on the right (i.e., purely optical) resonator \cite{LIT2009Guo,LITandEP2014Peng}. (b) The frequency spectrum of the compound COM system, with the red line or the blue lines denoting the red sideband (Stokes process) or the blue sidebands (anti-Stokes process), respectively \cite{OMIT2010Weis,OMIT2011Safavi}. }
\label{fig1}
\end{figure}

 As shown in Fig.\,\ref{fig1}, we consider two whispering-gallery-mode (WGM) microtoroid resonators coupled through evanescent fields, with the tunable coupling strength $J$ and the intrinsic optical loss $\gamma_1$ or $\gamma_2$, respectively \cite{PTdevice2014Peng,LITandEP2014Peng}. The external lights are input and output via tapered-fiber waveguides. As in Refs.\,\cite{LIT2009Guo,LITandEP2014Peng}, an additional optical loss $\gamma_{\text{tip}}$ is induced on the right (i.e., purely optical) resonator by a chromium (Cr)-coated silica nanofiber tip \cite{Tip2010Zhu}, in order to see the loss effects on such a compound COM system. The left resonator, supporting also a mechanical mode of frequency $\omega_{m}$ and an effective mass $m$, is driven by a strong red-detuned pump laser at frequency $\omega_L$ and a weak probe laser at frequency $\omega_P$ \cite{OMIT2010Weis,OMIT2011Safavi}, with the optical field amplitudes
\begin{equation}
\varepsilon_{L}=\sqrt{2\gamma_c P_L/\hbar\omega_L},~~~\varepsilon_{P}=\sqrt{2\gamma_c P_{\text{in}}/\hbar\omega_P},
\end{equation}
  respectively, where for simplicity we take $\gamma_1=\gamma_2=\gamma_c$ and $P_L$ or $P_{\text{in}}$ is the power of the pump or the probe light.

In a frame rotating at frequency $\omega_L$, the Hamiltonian of this compound COM system can be written at the simplest level as
\begin{eqnarray}
 \begin{aligned}
  H&=H_{0}+H_{\text{int}}+H_{\text{dr}},\\
H_{0}&=\frac{p^{2}}{2m}+\frac{1}{2}m\omega_{m}^{2}x^{2}+\hbar\Delta_{L}(a_{1}^{\dagger}a_{1}+a_{2}^{\dagger}a_{2}),\\
H_{\text{int}}&=-\hbar J(a_{1}^{\dagger}a_{2}+a_{2}^{\dagger}a_{1})-\hbar ga_{1}^{\dagger}a_{1}x,\\
H_{\text{dr}}&=i\hbar\varepsilon_{L}(a_{1}^{\dagger}-a_{1})+i\hbar\varepsilon_{P}(a_{1}^{\dagger}e^{-i\epsilon t}-a_{1}e^{i\epsilon t}),
 \label{Eq:Hamiltonian2}
 \end{aligned}
\end{eqnarray}
where $\omega_c$ is the resonant frequency of the optical  mode, $a_{1}$ ($a_{1}^{\dagger}$) and $a_{2}$($a_{2}^{\dagger}$)
are the optical bosonic annihilation (creation) operators, $g$ denotes the COM coupling strength, $x$ or $p$ is the mechanical displacement or momentum operator, and the optical detunings are
\begin{align}
\Delta_{L}=\omega_{c}-\omega_{L},~~~\epsilon=\omega_{P}-\omega_{L}.
\end{align}

The Heisenberg equations of motion of this compound system are
\begin{eqnarray}
\begin{aligned}
\ddot{x}&=-\Gamma_{\text{m}}\dot{x}-\omega_{m}^{2}x+\frac{\hbar g}{m}a_{1}^{\dagger}a_{1},\\
\dot{a}_{1}& =(-i\Delta_{L}-\gamma_{1}+igx)a_{1}+iJa_{2}+\varepsilon_{L}+\varepsilon_{P}e^{-i\epsilon t},\\
\dot{a}_{2}&=(-i\Delta_{L}-\gamma_{2}-\gamma_{\text{tip}})a_{2}+iJa_{1},
\label{Eq:equation of motion}
\end{aligned}
\end{eqnarray}
where $\Gamma_{\text{m}}$ is the mechanical loss rate. For $\varepsilon_{P} \ll \varepsilon_{L}$, we can take the probe light as a perturbation, the dynamical variables can be expressed as $a_{i}=a_{i,s}+\delta a_i$ ($i=$1,2) and $x=x_s+\delta x$, where the steady-state solutions of the system, by setting all the derivatives of the variables as zero, are easily obtained as
\begin{align}
\begin{aligned}
x_{s}&=\frac{\hbar g}{m\omega_{m}^{2}}\left|a_{1,s}\right|^{2},\\
a_{1,s}&=\frac{\varepsilon_{L}\left(i\Delta_{L}+\gamma_{2}+\gamma_{\text{tip}}\right)}{\left(i\Delta_{L}+\gamma_{1}-igx_{s}\right)\left(i\Delta_{L}+\gamma_{2}+\gamma_{\text{tip}}\right)+J^{2}},\label{as1}\\ a_{2,s}&=\frac{iJ\varepsilon_{L}}{\left(i\Delta_{L}+\gamma_{1}-igx_{s}\right)\left(i\Delta_{L}+\gamma_{2}+\gamma_{\text{tip}}\right)+J^{2}}.
\end{aligned}
\end{align}

\begin{figure}[htbp]
\centering
\includegraphics[width=12.5 cm]{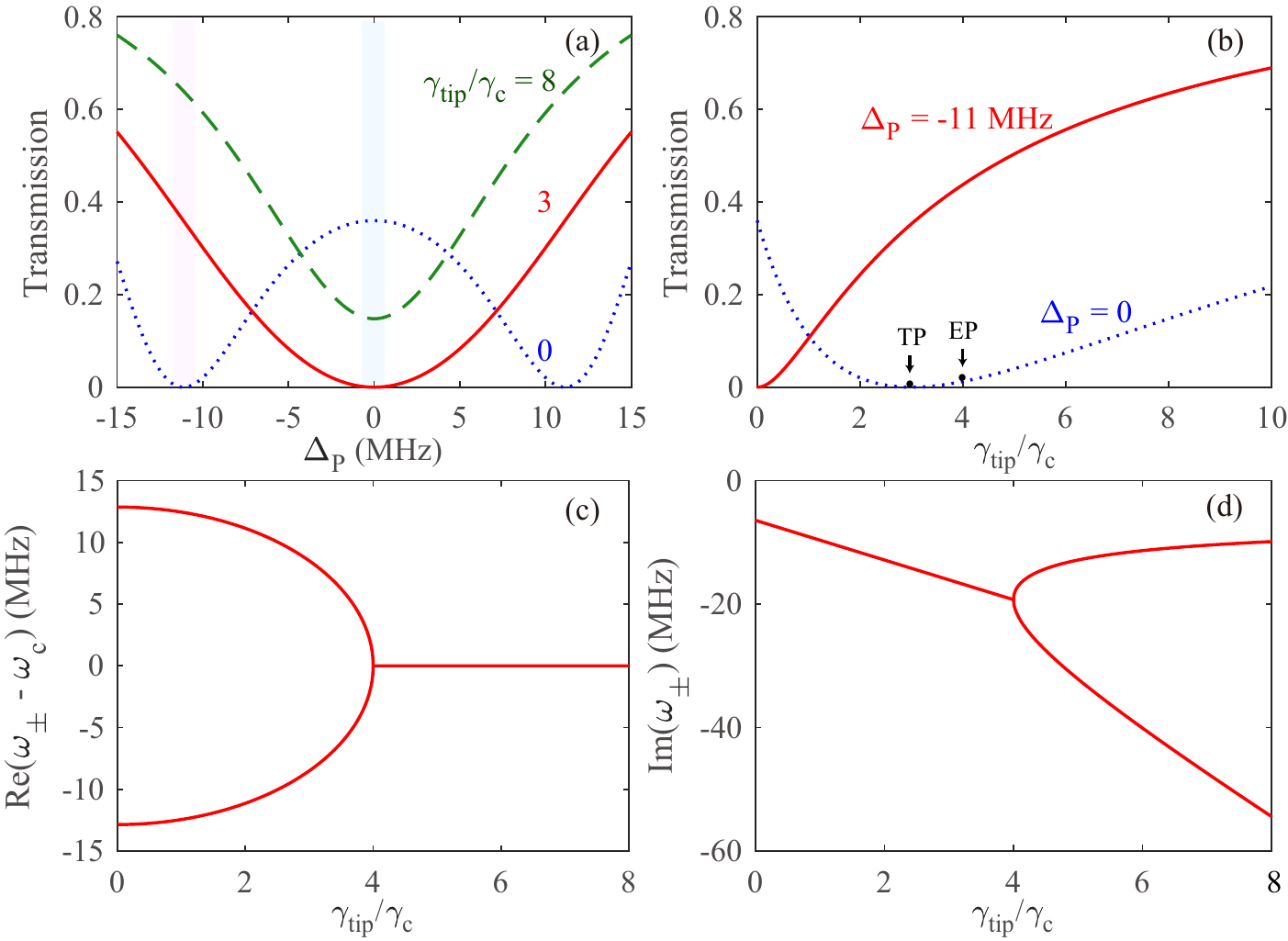}
\caption{(a) Transmission rate of the coupled optical system as a function of $\Delta_{\text{P}}$ at three selected $\gamma_{\text{tip}}/\gamma_c$.  (b) Transmission rate as a function of $\gamma_{\text{tip}}$ when $\Delta_{\text{P}}$ is 0 and -11 MHz. Evolution of the real (c) and imaginary (d) parts of the eigenfrequencies of the supermodes as a function of the loss  $\gamma_{\text{tip}}$. The parameters used here are $\gamma_1=\gamma_2=\gamma_c=6.43$ MHz and $J/\gamma_c=2$.}
\label{fig2}
\end{figure}

For comparisons, we first consider the purely optical case \cite{LIT2009Guo} by ignoring the COM coupling.
In this special case, by using the input-output relation \cite{IOR1985Gardiner} $
a^{\text{out}}_1=a^{\text{in}}_1-\sqrt{2\gamma_{1}}a_1$, we can derive the optical transmission rate as
\begin{align}
T =\left|\frac{a_{1}^{\text{out}}}{a_{1}^{\text{in}}}\right|^{2} =\left|1-\frac{2\gamma_{1}\left(i\Delta_{2}+\gamma_{2}+\gamma_{\text{tip}}\right)}{\left(i\Delta_{1}+\gamma_{1}\right)\left(i\Delta_{2}+
\gamma_{2}+\gamma_{\text{tip}}\right)+J^{2}}\right|^{2},
\label{CRITT}
\end{align}
where $\Delta_i=\omega_P - \omega_i$ ($i=1,~2$) is the detuning between the probe and the cavity mode. For simplicity, here we take $\Delta_\text{1}=\Delta_\text{2}=\Delta_\text{P}$. As shown in Fig.\,\ref{fig2}(a), the LIT feature can be clearly seen at the resonance (i.e., $\Delta_{\text{P}}=0$) in the transmission spectrum, that is, the transmission rate firstly decreases and then increases by increasing the tip loss $\gamma_{\text{tip}}$ \cite{LIT2009Guo}. The turning point (TP) position turns out to be
\begin{equation}
\gamma^{\text{TP}}_{\text{tip}}=-(i\Delta_2 + \gamma_2) + \frac{(i\Delta_1 + \gamma_1) J^2}{\Delta_1^2 + \gamma_1^2},
\end{equation}
which, for the parameter values chosen here, corresponds to $\gamma^{\text{TP}}_{\text{tip}}/\gamma_c=3$ (illustrated in Fig.\,\ref{fig2}(b)).
Interestingly, we also note that by increasing $\gamma_{\text{tip}}$, the strong-absorption regimes in the conventional transmission spectrum (at $\Delta_{\text{P}}=\pm 11$ MHz) become transparent, see Fig.\,\ref{fig2}(b), which is not reported in Ref.\,\cite{LIT2009Guo}. This phenomenon is induced by the reduction of interference caused by the tip loss. And from our numerical estimation, $\gamma^{\text{TP}}_{\text{tip}}$ is $\sim0$ for $\Delta_{\text{P}}=\pm 11$ MHz, i.e., by reversing the tip loss to an active gain, it is possible to reverse the dip in the EIT spectrum to a peak as already observed in the recent reversed EIT experiment performed by T. Oishi and M. Tomita \cite{ReversedEIT}.
LIT is generally viewed as the evidence of the EP emergence in this lossy system \cite{LIT2009Guo}, or the existence of hidden parity-time symmetry (under a suitable mathematical transformation) \cite{PTReview2014Zyablovsky}. The eigenfrequencies of this coupled optical system are
\begin{align}
\omega_{\pm}=\frac{1}{2}\left[\left(\omega_{1}+\omega_{2}\right)
-i\left(\gamma_{1}+\gamma_{2}+\gamma_{\text{tip}}\right)\right]\pm\frac{1}{2}
\sqrt{\left[\left(\omega_{1}-\omega_{2}\right)+i\left(\gamma_{2}
+\gamma_{\text{tip}}-\gamma_{1}\right)\right]^{2}+4J^{2}}.
\end{align}
For $\omega_{1}=\omega_{2}=\omega_{c}$, the EP condition is simplified as $\gamma_{\text{tip}}^{\mathrm{EP}} = \gamma_{1} - \gamma_{2} + 2J$, or for the parameter values chosen here, $\gamma_{\text{tip}}^{\mathrm{EP}}/\gamma_c=4$, see Fig.\,\ref{fig2}(c-d). Clearly, $\gamma^{\text{TP}}_{\text{tip}}$ can be close to but not exactly coincides with $\gamma_{\text{tip}}^{\mathrm{EP}}$, due to the fact that the TP depends on the detuning $\Delta_{1,2}$ while the EP does not (for similar features, see also Ref.\,\cite{LITandEP2014Peng}).

\begin{figure}[htb!]
\centering
\includegraphics[width=13 cm]{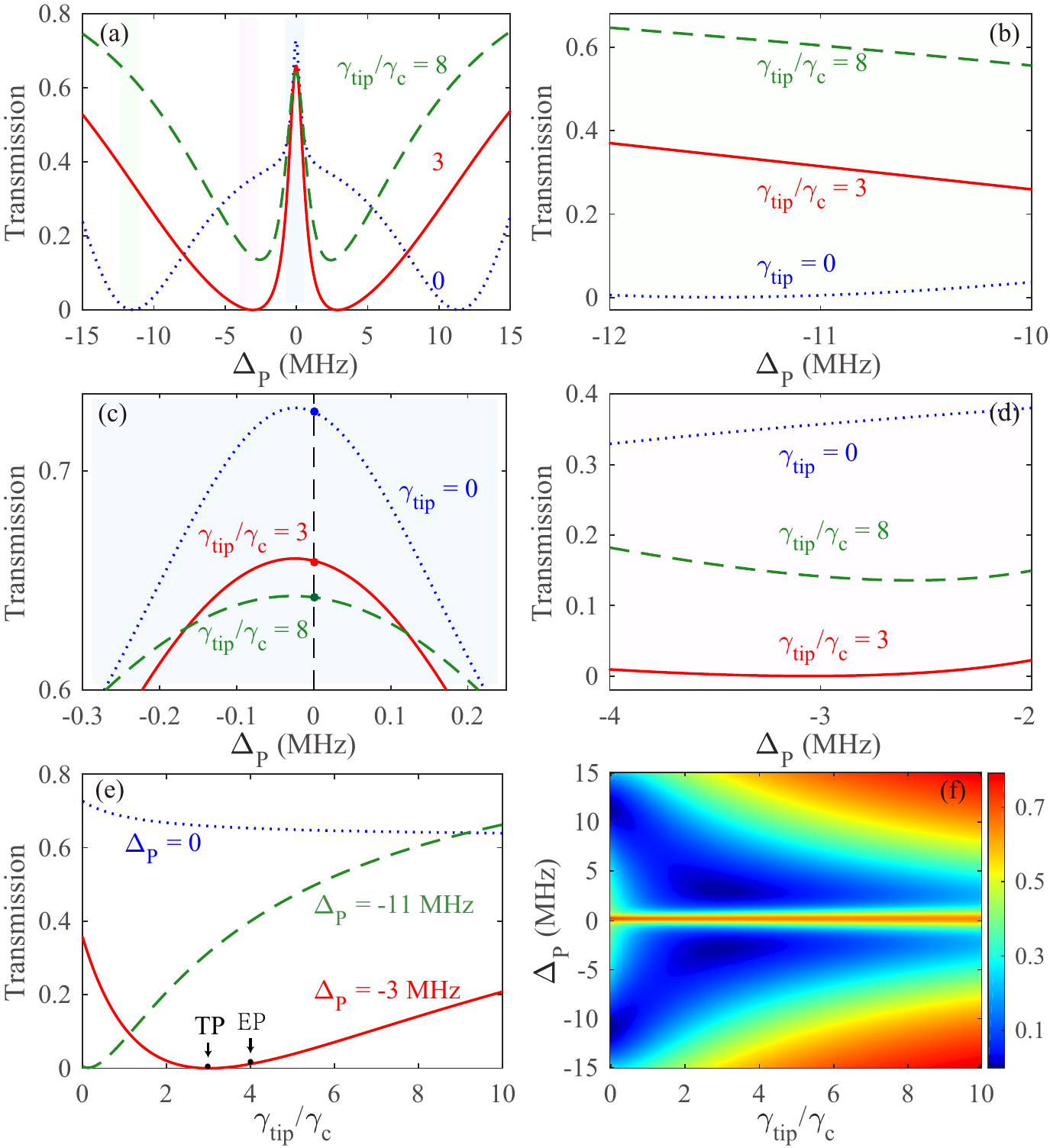}
\caption{(a)--(d) Transmission rate $T_\text{P}$ of the probe light as a function of $\Delta_\text{P}$. (e) Transmission rate of OMIT as a function of $\gamma_{\text{tip}}$ at different $\Delta_{\text{P}}$. (f) Transmission rate of OMIT as a function of $\gamma_{\text{tip}}$ and $\Delta_{\text{P}}$. The other parameters are $\omega_{c}=193$ THz, $\gamma_{c}=6.43$ MHz, $\omega_{m}=2\pi\times23.4$ MHz, $P_{L}=1$ mW, $\Delta_{L}=\omega_{m}$, $g=\omega_{c}/R$, $R=34.5\ \mu\text{m}$, $m=5\times10^{-11}$ kg, $J=12.86$ MHz and $\Gamma_{\text{m}}=0.24$
MHz.}
\label{fig3}
\end{figure}

Now we consider the role of COM coupling in LIT. For this aim, we express the dynamical variables as the sum of their steady-state values and small fluctuations to the first order, i.e.,
\begin{align}
x=x_{s}+\delta x^{(1)}+\cdot\cdot\cdot,~~~
a_{i}=a_{i,s}+\delta a_{i}^{(1)}+\cdot\cdot\cdot\quad (i=1,2),
\end{align}
with which we can rewrite the equations of motion as
\begin{align}
\begin{aligned}
\frac{d^{2}}{dt^{2}}(x_{s}+\delta x^{(1)})&=-\Gamma_{\text{m}}\frac{d}{dt}(x_{s}+\delta x^{(1)})-\omega_{m}^{2}(x_{s}+\delta x^{(1)})+\frac{\hbar g}{m}(a_{1,s}+\delta a_{1}^{(1)})(a_{1,s}+\delta a_{1}^{(1)}),\\
\frac{d}{dt}(a_{1,s}+\delta a_{1}^{(1)}) &=[-i\Delta_{L}-\gamma_{1}+ig(x_{s}+\delta x^{(1)})](a_{1,s}+\delta a_{1}^{(1)})+iJ(a_{2,s}+\delta a_{2}^{(1)})\\
&\quad +\varepsilon_{L}+\varepsilon_{P}e^{-i\epsilon t},\\
\frac{d}{dt}(a_{2,s}+\delta a_{2}^{(1)}) &=(-i\Delta_{L}-\gamma_{2}-\gamma_{\text{tip}})(a_{2,s}+\delta a_{2}^{(1)})+iJ(a_{1,s}+\delta a_{1}^{(1)}).
\end{aligned}
\end{align}
Here the higher-order terms such as $\delta x^{(1)}\delta a_{1}^{(1)}$ will be neglected since they only contribute to the higher-order sidebands \cite{HigherOrderOMIT2012Xiong}.

Then by using the ansatz:
\begin{align}
\left(\begin{array}{c}
\text{\ensuremath{\delta}}a_{i}^{(1)}\\
\delta x^{(1)}
\end{array}\right)&=\left(\begin{array}{c}
\text{\ensuremath{\delta}}a_{i+}^{(1)}\\
\delta x_{+}^{(1)}
\end{array}\right)e^{-i\epsilon t}+\left(\begin{array}{c}
\text{\ensuremath{\delta}}a_{i-}^{(1)}\\
\delta x_{-}^{(1)}
\end{array}\right)e^{i\epsilon t}\quad (i=1,2),
\label{Eq:FirstOrder}
\end{align}
we obtain the solutions for the fluctuation operators as
\begin{align}
\delta x_+^{(1)}&=\frac{\hbar g\varepsilon_{P}a_{1,s}^{\ast}\mu_{-}^{(1)}\mathcal{A}_{1}^{(1)}}{\mathcal{K}^{(1)}\mathcal{A}_{1}^{(1)}\mathcal{A}_{2}^{(1)}+i\hbar g^{2}\left|a_{1,s}\right|^{2}\left(\mu_{+}^{(1)\ast}\mathcal{A}_{2}^{(1)}-\mu_{-}^{(1)}\mathcal{A}_{1}^{(1)}\right)},\\
\delta a_{1+}^{(1)}&=\frac{\varepsilon_{P}\mu_{-}^{(1)}\left(\mathcal{K}^{(1)}\mathcal{A}_{1}^{(1)}+i\hbar g^{2}\left|a_{1,s}\right|^{2}\mu_{+}^{(1)\ast}\right)}{\mathcal{K}^{(1)}\mathcal{A}_{1}^{(1)}\mathcal{A}_{2}^{(1)}+i\hbar g^{2}\left|a_{1,s}\right|^{2}\left(\mu_{+}^{(1)\ast}\mathcal{A}_{2}^{(1)}-\mu_{-}^{(1)}\mathcal{A}_{1}^{(1)}\right)},
\label{Eq:linear variable}
\end{align}
where $\mathcal{K}^{(1)}=m\left(-\epsilon^{2}-i\epsilon\Gamma_{\text{m}}+\omega_{m}^{2}\right)$ and
\begin{align}
\begin{array}{l}
  \mathcal{A}_{1}^{(1)}=\mu_{+}^{(1)\ast}\nu_{+}^{(1)\ast}+J^{2},\qquad \quad \ \  \mathcal{A}_{2}^{(1)}=\mu_{-}^{(1)}\nu_{-}^{(1)}+J^{2},\\
  \mu_{\pm}^{(1)}=i\Delta_{L}+\gamma_{2}+\gamma_{\text{tip}}\pm i\epsilon,\qquad \nu_{\pm}^{(1)}=i\Delta_{L}+\gamma_{1}-igx_{s}\pm i\epsilon.
\end{array}
\end{align}

With these results at hand, by using the standard input-output relation \cite{IOR1985Gardiner}, we can obtain the transmission rate of the probe light
\begin{align}
T_{\text{P}}=\left|\frac{a_{1}^{\text{out}}}{a_{1}^{\text{in}}}\right|^{2}=\left|\frac{\varepsilon_{P}-2\gamma_{1}\delta a_{1+}^{(1)}}{\varepsilon_{P}}\right|^{2}=\left|1-\frac{2\gamma_{1}\delta a_{1+}^{(1)}}{\varepsilon_{P}}\right|^{2},
\end{align}
which describes the relation of the output field amplitude and the input field amplitude at the probe frequency.

Figure\,\ref{fig3} shows the transmission rate $T_\text{P}$ of the probe as a function of $\Delta_{\text{P}}$ and $\gamma_{\text{tip}}$. We see that (i) the strong-absorption regimes at $\Delta_{\text{p}} = \pm 11\,\mathrm{MHz}$ become transparent by increasing the tip loss, e.g., $T_{\text{P}}$ is increased from zero to $\sim 0.35$ or $\sim 0.6$ for $\gamma_{\text{tip}}/\gamma_c\sim 3$ or $8$, see Fig.\,\ref{fig3}(b) which is same as the purely optical case. In contrast, (ii) for the resonant case ($\Delta_{\text{P}}=0$), the OMIT peak tends to be lowered, i.e., $T_{\text{P}}$ decreases for more tip loss, with its linewidth firstly decreased but then increased again, see Fig.\,\ref{fig3}(a,c). More interestingly, (iii) for the intermediate regime ($\Delta_{\text{P}} =\pm 3\,\mathrm{MHz}$), the feature as LIT in purely optical systems \cite{LIT2009Guo,LITandEP2014Peng} in the resonant case can be clearly seen, i.e., $T_{\text{P}}$ firstly drops down to zero but then increases again for more tip loss, with the turning point $\gamma^{\text{TP}}_{\text{tip}}/\gamma_c=3$, see Fig.\,\ref{fig3}(d).
A more intuitive analysis of these phenomena in the various parametric regimes is shown in Fig.\,\ref{fig3}(e) and the TP and EP are illustrated in the figure. In Fig.\,\ref{fig3}(f), the transmission rate is plotted as a function of $\gamma_{\text{tip}}$ and $\Delta_{\text{P}}$ which are both continuously varying to give a comprehensive view.

\begin{figure}[htbp]
  \centering
  \includegraphics[width=13 cm]{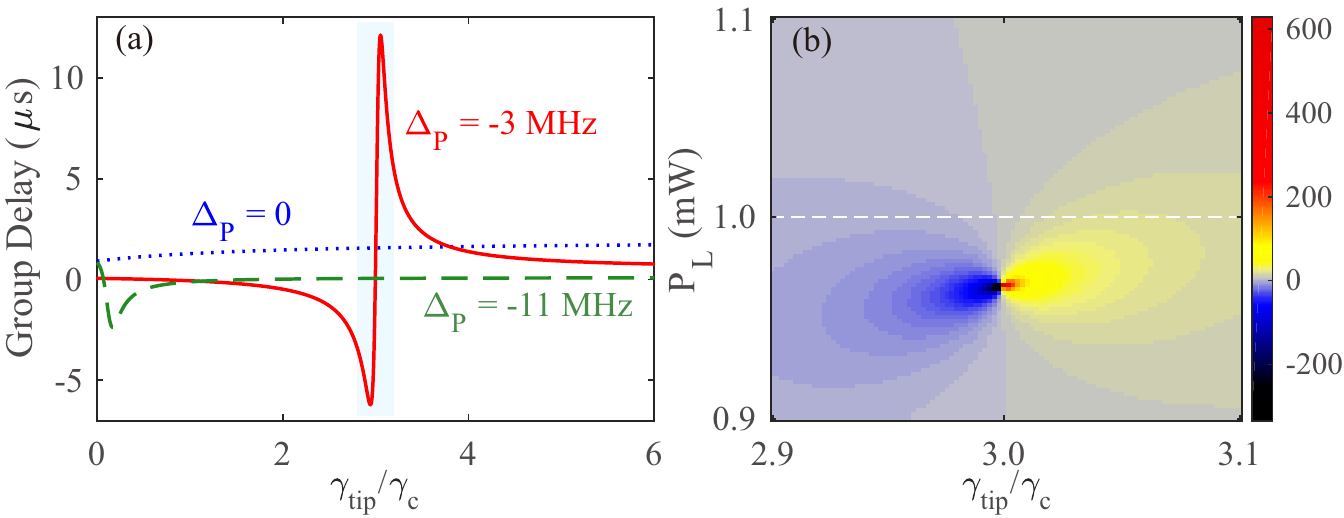}
  \caption{(a) Group delay of the probe light as a function of $\gamma_{\text{tip}}$ at different $\Delta_{\text{P}}$. The pump power $P_L$ is $1$ mW. (b) Group delay of the probe light as a function of $\gamma_{\text{tip}}$ and the pump power $P_L$ at $\Delta_{\text{P}}=-3$ MHz. The unit of group delay is $\mu s$.}\label{fig4}
\end{figure}

Based on the analyses above, we find that the LIT emerging at the resonance in the purely optical system now moves to the off-resonance regime of a specific detuning ($\Delta_{\text{P}}=\pm3$ MHz) in the COM system. We will give an analysis of this difference in the following section. By comparing the linearized equations of $\delta a^{(1)}_{1+}$ corresponding to the optical case and the COM case
\begin{align}
\left(i\Delta_{L}+\gamma_{1}-i\epsilon\right)\delta a^{(1)}_{1+}&=iJ\delta a^{(1)}_{2+}+\varepsilon_{P},\label{op}\\
\left(i\Delta_{L}+\gamma_{1}-igx_{s}-i\epsilon\right)\delta a^{(1)}_{1+}&=iga_{1,s}\delta x^{(1)}_{+}+iJ\delta a^{(1)}_{2+}+\varepsilon_{P},\label{om}
\end{align}
we can see that the differences are the COM terms: $-igx_{s}\delta a^{(1)}_{1+}\text,\ \ iga_{1,s}\delta x^{(1)}_{+}$. Then we transform Eq.\,(\ref{om}) into a form similar to Eq.\,(\ref{op}) utilizing the linearized equation of $\delta x^{(1)}_{+}$
\begin{align}
\left(i\Delta^{\prime}+\gamma^{\prime}_{1}-i\epsilon\right)\delta a^{(1)}_{1+}=iJ\delta a^{(1)}_{2+}+\varepsilon_{P},
\end{align}
where
\begin{equation}
\Delta^{\prime}=\Delta_{L}-gx_{s}-\textbf{Re}(C_{1}),~~~\gamma^{\prime}_{1}=\gamma_{1}+\textbf{Im}(C_{1}),
\end{equation}
and
\begin{align}
C_{1}&=\frac{\mathcal{A}_{1}^{(1)}\hbar g^{2}\left|a_{1,s}\right|^{2}}{\mathcal{A}_{1}^{(1)}\mathcal{K}^{(1)}+i\hbar g^{2}\left|a_{1,s}\right|^{2}\left(-i\Delta_{L}+\gamma_{2}+\gamma_{\text{tip}}-i\epsilon\right)}.
\end{align}
With chosen parameters above, we can numerically estimate a frequency shift of $gx_{s}+\textbf{Re}(C_{1})$ to be $\sim3$ MHz in the steady-state case. For comparison with purely optical system, we choose $\Delta_{L}=0$ and $\Delta_{\text{P}}=0$. This frequency shift caused by the COM interaction is matched well with the parametric regimes where we find the LIT at the detuning of $\Delta_{\text{P}}=\pm3$ MHz in the transmission spectrum of OMIT.

Now we turn to the slow-to-fast-light switch at the EP. Slowing or advancing of light can be associated with the OMIT process due to the abnormal dispersion \cite{OMIT2011Safavi}. This feature can be characterized by the group delay of the probe light
\begin{align}
\tau_{g}=\frac{\text{d}\,\text{arg}(T_{\text{P}})}{\text{d}\,\Delta_{\text{P}}},
\label{gd}
\end{align}

Figure\,\ref{fig4} shows the group delay as a function of $\gamma_{\text{tip}}$ at different values of $\Delta_{\text{P}}$. We find that at $\Delta_{\text{P}} = -3$ MHz, the probe light experiences a fast-to-slow switch in the vicinity of the EP, a feature which is similar to the reverted OMIT reported previously in an active COM system \cite{PTOMIT2015Jing}. This provides a new method to achieve coherent optical group-velocity switch by tuning the optical loss, which as far as we know, has not been demonstrated previously in purely optical systems. In view of the sensitive change of $\tau_g$ at the EP, this also could be used for e.g., EP-enhanced sensing of external particles entering into the mode volume of the resonator \cite{EPSensing2017Hodaei,EPSensing2017Chen,EPSensor2018Hassan}.

\begin{figure}[htbp]
\centering
\includegraphics[width=13.1 cm]{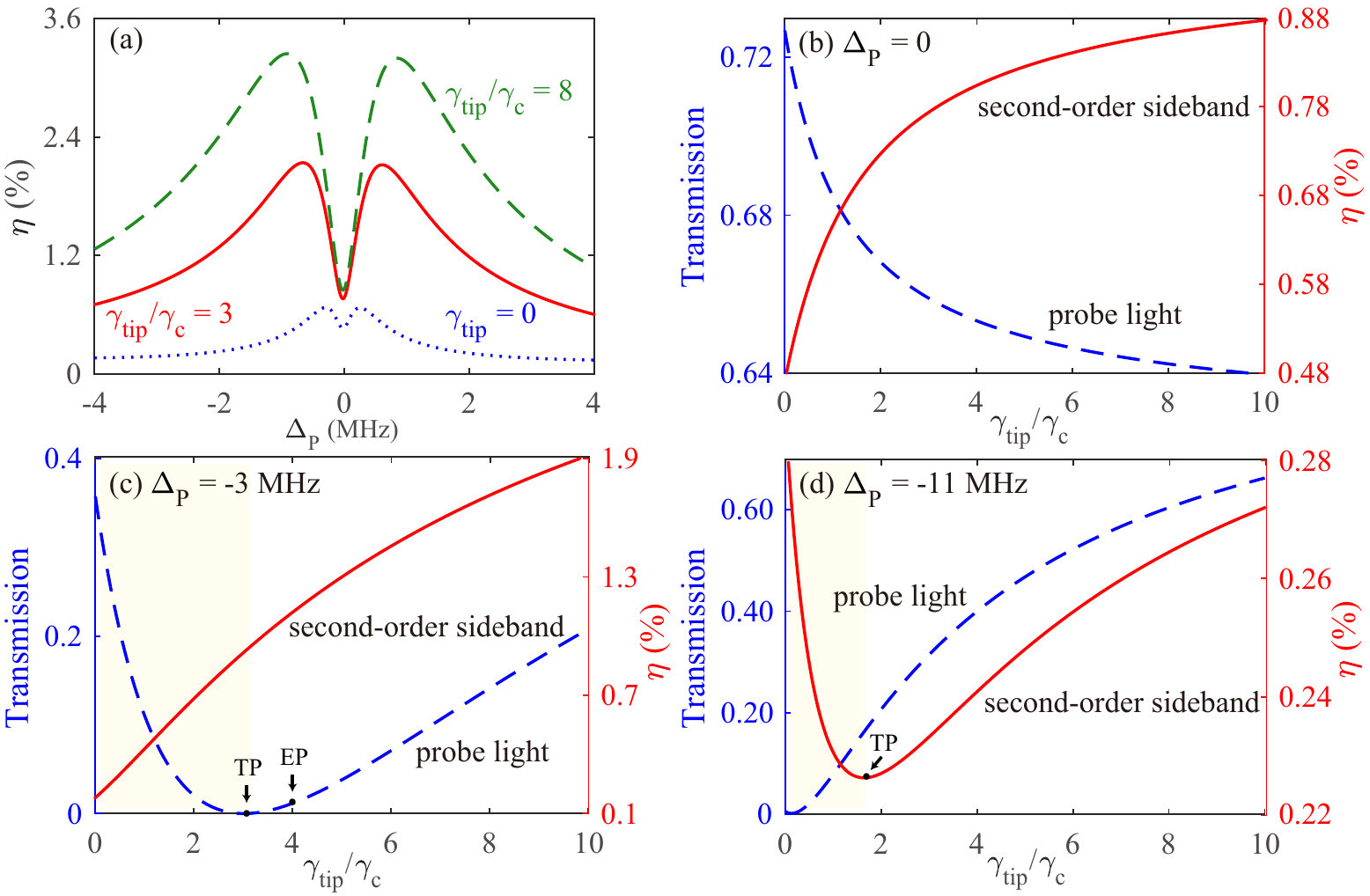}
\caption{(a) The efficiency of the second-order sideband process as a function of $\Delta_\text{P}$. Subfigures (b), (c) and (d) show the comparisons between transmission rate of OMIT and efficiency of the second-order sideband as a function of $\gamma_{\text{tip}}$.}
\label{fig5}
\end{figure}

\section{Second-order LIT in COM}

In contrast to the linear systems \cite{LIT2009Guo,LITandEP2014Peng}, the tip loss can also affect the higher-order process originating from intrinsic nonlinear COM interactions. In order to see this, we use the following second-order expansions of the operators \cite{HigherOrderOMIT2012Xiong}
\begin{align}
x=x_{s}+\delta x^{(1)}+\delta x^{(2)}+\cdot\cdot\cdot,~~~a_{i}=a_{i,s}+\delta a_{i}^{(1)}+\delta a_{i}^{(2)}+\cdot\cdot\cdot\quad (i=1,\,2),
\label{Eq:ESecondOrder}
\end{align}
with
\begin{align}
\left(\begin{array}{c}
\text{\ensuremath{\delta}}a_{i}^{(2)}\\
\delta x^{(2)}
\end{array}\right)&=\left(\begin{array}{c}
\text{\ensuremath{\delta}}a_{i+}^{(2)}\\
\delta x_{+}^{(2)}
\end{array}\right)e^{-2i\epsilon t}+\left(\begin{array}{c}
\text{\ensuremath{\delta}}a_{i-}^{(2)}\\
\delta x_{-}^{(2)}
\end{array}\right)e^{2i\epsilon t}\quad (i=1,2).
\label{Eq:SecondOrder}
\end{align}
Then by solving the Eq.\,(\ref{Eq:equation of motion}), with the aid of Eqs.\,(\ref{Eq:FirstOrder},\ref{Eq:ESecondOrder},\ref{Eq:SecondOrder}) and neglecting the higher-order terms more than second order, we get the second-order solutions
\begin{align}
\text{\ensuremath{\delta}}a_{1+}^{(2)}=\frac{-i\hbar g^{4}a_{1,s}\left|a_{1,s}\right|^{2}\mu_{+}^{(1)\ast}\mu_{+}^{(2)\ast}\mu_{-}^{(2)}\delta x_{+}^{(1)2}+\lambda\delta x_{+}^{(1)}\text{\ensuremath{\delta}}a_{1+}^{(1)}}{\mathcal{A}_{1}^{(1)}\left[\mathcal{K}^{(2)}\mathcal{A}_{1}^{(2)}\mathcal{A}_{2}^{(2)}+i\hbar g^{2}\left|a_{1,s}\right|^{2}\left(\mathcal{A}_{2}^{(2)}\mu_{+}^{(2)\ast}-\mathcal{A}_{1}^{(2)}\mu_{-}^{(2)}\right)\right]},
\end{align}
with $\mathcal{K}^{(2)}=m(-4\epsilon^{2}-2i\epsilon\Gamma_{\text{m}}+\omega_{m}^{2})$ and
\begin{align}
\begin{array}{l}
\mathcal{A}_{1}^{(2)}=\mu_{+}^{(2)\ast}\nu_{+}^{(2)\ast}+J^{2},\qquad \qquad \
\mathcal{A}_{2}^{(2)}=\mu_{-}^{(2)}\nu_{-}^{(2)}+J^{2},\\
\mu_{\pm}^{(2)}=i\Delta_{L}+\gamma_{2}+\gamma_{\text{tip}}\pm 2i\epsilon,\qquad \nu_{\pm}^{(2)}=i\Delta_{L}+\gamma_{1}-igx_{s}\pm 2i\epsilon, \\
\lambda=ig\mu_{-}^{(2)}\mathcal{K}^{(2)}\mathcal{A}_{1}^{(1)}\mathcal{A}_{1}^{(2)}-\hbar g^{3}\left|a_{1,s}\right|^{2}\mu_{-}^{(2)}\left(\mathcal{A}_{1}^{(1)}\mu_{+}^{(2)\ast}-\mu_{+}^{(1)\ast}\mathcal{A}_{1}^{(2)}\right).
\end{array}
\end{align}

As defined in Ref.\cite{HigherOrderOMIT2012Xiong}, the efficiency of the second-order sideband process is
\begin{align}
\eta=\left|-\frac{2\gamma_{1}}{\varepsilon_{P}}\delta a_{1+}^{(2)}\right|.
\end{align}
Figure \ref{fig5}(a) shows the impact of the tip loss on the second-order sideband of OMIT. We find that, in contrast to the linear cases, the efficiency $\eta$ increases by enhancing the loss $\gamma_{\text{tip}}$. A comparison between $T_{\text{P}}$ and $\eta$ is shown in Fig.\,\ref{fig5}(b-d). Figure \ref{fig5}(b) shows that $T_{\text{P}}$ decreases by increasing $\gamma_{\text{tip}}$ at the resonance while $\eta$ is enhanced, a feature which was firstly revealed in Ref.\,\cite{HigherOrderOMIT2012Xiong}. However, for non-resonance cases, e.g., $\Delta_{\text{P}}=\pm3\,$MHz or $\Delta_{\text{P}}=\pm11\,$MHz, $\eta$ increases by increasing $\gamma_{\text{tip}}$, which is evidently different from that for $T_{\text{P}}$, see Fig.\,\ref{fig5}(c-d). Clearly, these results on nonlinear OMIT process are beyond any linear EP picture \cite{HighOrderEPPhononCooling2017Jing}. We note that the presence of nonlinearity can lead to a shift of the EP position \cite{NLEP2012Konotop} or even the emergence of high-order EPs \cite{HighOrderEPPhononCooling2017Jing}.
We also note that $\eta_\mathrm{TP}$ emerges at $\Delta_{\text{P}}=\pm11\,$MHz, which is clearly different from the linear TP occurring at $\Delta_{\text{P}}=\pm3\,$MHz. In fact, this frequency shift can also be identified by comparing the equations describing the linear process and its second-order sidebands, i.e.,
\begin{align}
\left(i\Delta^{\prime}+\gamma^{\prime}_{1}-i\epsilon\right)\delta a^{(1)}_{1+} & =iJ\delta a^{(1)}_{2+}+\varepsilon_{P},\\
\left(i\Delta^{\prime\prime}+\gamma^{\prime\prime}_{1}-2i\epsilon\right)\delta a^{(2)}_{1+} & =iJ\delta a^{(2)}_{2+}+\mathcal{B},
\end{align}
with
\begin{equation}
\Delta^{\prime\prime}=\Delta_{L}-gx_{s}-\textbf{Re}(C_{2}),~~~\gamma^{\prime\prime}_{1}=\gamma_{1}+\textbf{Im}(C_{2}),
\end{equation}
and
\begin{align}
  C_{2} &= \frac{\hbar g^{2}\left|a_{1,s}\right|^{2}\mathcal{A}_{1}^{(2)}\mathcal{A}_{1}^{(1)}}{\mathcal{A}_{1}^{(2)}\mathcal{A}_{1}^{(1)}\mathcal{K}^{(2)}+i\hbar g^{2}|a_{1,s}|^{2}\mathcal{A}_{1}^{(1)}\mu_{+}^{(2)\ast}},\\
  \mathcal{B} &= \frac{i\hbar g^{2}a_{1,s}\left[-g^{2}|a_{1,s}|^{2}\mu_{+}^{(2)\ast}\mu_{+}^{(1)\ast}\delta x_{+}^{(1)2}-ig\mathcal{A}_{1}^{(2)}a_{1,s}^{\ast}\mu_{+}^{(1)\ast}\delta x_{+}^{(1)}\text{\ensuremath{\delta}}a_{1+}^{(1)}\right]}{\mathcal{A}_{1}^{(2)}\mathcal{A}_{1}^{(1)}\mathcal{K}^{(2)}+i\hbar g^{2}\mathcal{A}_{1}^{(1)}|a_{1,s}|^{2}\mu_{+}^{(2)\ast}}+ig\delta x_{+}^{(1)}\text{\ensuremath{\delta}}a_{1+}^{(1)}.
\end{align}
Here $\mathcal{B}$, in terms of $\delta x_{+}^{(1)}$ and $\text{\ensuremath{\delta}}a_{1+}^{(1)}$, can be taken as a constant. We see that in comparison with the linear process, the second-order sideband experiences a frequency shift $|\textbf{Re}(C_{2})-\textbf{Re}(C_{1})|\sim10\,$MHz, i.e., agreeing with our numerical results in the order of magnitudes.

\section{Conclusions and Discussions}

We theoretically investigate the impact of loss on OMIT in a passive compound COM system by coupling an external nanotip to the optical resonator. Loss-induced transparency is found at the EP in the OMIT transmission spectrum, which is reminiscent of that as reported in Ref.\,\cite{LIT2009Guo}, but here corresponding to the off-resonance case (i.e., with $\Delta_{\text{P}}=\pm3$ MHz). For the resonance case, however, increasing the tip loss leads to very minor changes for the OMIT peak. We also find that a slow-to-fast light switch can happen in the vicinity of the loss-induced EP. A detailed comparison between the linear OMIT process and its second-order sidebands, in the presence of a tunable tip loss, is also given, indicating that more exotic EP-assisted effects may happen in a nonlinear COM system. In practice, our work provides a promising new way to manipulate or switch both light transmissions and optical group delays with various COM devices. Finally, in view of the sensitive change of the optical group delay at the EP, our work also indicates a new way to achieve EP-enhanced sensing \cite{EPSensing2017Hodaei,EPSensing2017Chen,EPSensor2018Hassan}.

\section*{Funding}
This work is supported by NSF of China under Grants No. 11474087 and No.
11774086, and the HuNU Program for Talented Youth.


\begin{thebibliography}{99}
\bibitem{Aspelmeyer2014CavityOptomechanics}M. Aspelmeyer, T. J. Kippenberg, and F. Marquardt, "Cavity optomechanics," Rev. Mod. Phys. {\bfseries 86}(4), 1391--1452 (2014).
\bibitem{Metcalfe2014Applications}M. Metcalfe, "Applications of cavity optomechanics," App. Phys. Rev. {\bfseries 1}(3), 031105 (2014).
\bibitem{Milestone2006}Milestones: Photons, \url{https://www.nature.com/milestones/milephotons/timeline.html}.
\bibitem{Optomechanicalcavity2008Kippenberg}T. J. Kippenberg and K. J. Vahala, "Cavity optomechanics: back-action at the mesoscale," Science {\bfseries 321}(5893), 1172--1176 (2008).
\bibitem{Atomicgas2008Brennecke}F. Brennecke, S. Ritter, T. Donner, and T. Esslinger, "Cavity optomechanics with a Bose-Einstein condensate," Science {\bfseries 322}(5899), 235--238 (2008).
\bibitem{Droplet2016Dahan}R. Dahan, L. L. Martin, and T. Carmon, "Droplet optomechanics," Optica {\bfseries 3}(2), 175--178 (2016).
\bibitem{Qt2010Rabl}P. Rabl, S. J. Kolkowitz, F. H. L. Koppens, J. G. E. Harris, P. Zoller, and M. D. Lukin, "A quantum spin transducer based on nanoelectromechanical resonator arrays," Nat. Phys. {\bfseries 6}(8), 602--608 (2010).
\bibitem{Qt2013Bochmann}J. Bochmann, A. Vainsencher, D. D. Awschalom, and A. N. Cleland, "Nanomechanical coupling between microwave and optical photons," Nat. Phys. {\bfseries 9}(11), 712--716 (2013).
\bibitem{QtOptoe2018Midolo}L. Midolo, A. Schliesser, and A. Fiore, "Nano-opto-electro-mechanical systems," Nat. Nanotechnol. {\bfseries 13}(1), 11--18 (2018).
\bibitem{Squeezing2015Wollman}E. E. Wollman, C. U. Lei, A. J. Weinstein, J. Suh, A. Kronwald, F. Marquardt, A. A. Clerk, and K. C. Schwab, "Quantum squeezing of motion in a mechanical resonator," Science {\bfseries 349}(6251), 952--955 (2015).
\bibitem{PhononLaser2010Grudinin}I. S. Grudinin, H. Lee, O. Painter and K. J. Vahala, "Phonon laser action in a tunable two-level system," Phys. Rev. Lett. {\bfseries 104}(8), 083901 (2010).
\bibitem{PhononLaser2014Jing}H. Jing, \c{S}. K. \"{O}zdemir, X. Y. L\"{u}, J. Zhang, L. Yang, and F. Nori, "$\mathscr{PT}$-symmetric phonon laser," Phys. Rev. Lett. {\bfseries 113}(5), 053604 (2014).
\bibitem{EPPhononLaser2017Lv}H. L\"{u}, \c{S}. K. \"{O}zdemir, L. M. Kuang, F. Nori, and H. Jing, "Exceptional points in random-defect phonon lasers," Phys. Rev. Appl. {\bfseries 8}(4), 044020 (2017).
\bibitem{Sensing2010Krause}A. G. Krause, M. Winger, T. D. Blasius, Q. Lin, and O. Painter, "A high-resolution microchip optomechanical accelerometer," Nat. Photonics {\bfseries 6}(11), 768--772 (2012).
\bibitem{Measurement2012Gavartin}E. Gavartin, P. Verlot, and T. J. Kippenberg, "A hybrid on-chip optomechanical transducer for ultrasensitive force measurements," Nat. Nanotechnol. {\bfseries 7}(8), 509--514 (2012).
\bibitem{OMIT2010Agarwal}G. S. Agarwal and S. Huang, "Electromagnetically induced transparency in mechanical effects of light," Phys. Rev. A {\bfseries 81}(4), 041803 (2010).
\bibitem{RAonOMIT2017LiuYongChun}Y.-C. Liu, B.-B. Li, and Y.-F. Xiao, "Electromagnetically induced transparency in optical microcavities," Nanophotonics {\bfseries 6}(5), 789--811 (2017).
\bibitem{OMIT2010Weis}S. Weis, R. Rivi\`{e}re, S. Del\'{e}glise, E. Gavartin, O. Arcizet, A. Schliesser, and T. J. Kippenberg, "Optomechanically induced transparency," Science {\bfseries 330}(6010), 1520--1523 (2010).
\bibitem{OMIT2011Safavi}A. H. Safavi-Naeini, T. P. Mayer Alegre, J. Chan, M. Eichenfield, M. Winger, Q. Lin, J. T. Hill, D. E. Chang, and O. Painter, "Electromagnetically induced transparency and slow light with optomechanics," Nature {\bfseries 472}(7341), 69--73 (2011).
\bibitem{OMIT2011Teufel}J. D. Teufel, D. Li, M. S. Allman, K. Cicak, A. J. Sirois, J. D. Whittaker, and R. W. Simmonds, "Circuit cavity electromechanics in the strong-coupling regime," Nature {\bfseries 471}(7337), 204--208 (2011).
\bibitem{OMIT2013Karuza}M. Karuza, C. Biancofiore, M. Bawaj, C. Molinelli, M. Galassi, R. Natali, P. Tombesi, G. Di Giuseppe, and D. Vitali, "Optomechanically induced transparency in a membrane-in-the-middle setup at room temperature," Phys. Rev. A {\bfseries 88}(1), 013804 (2013).
\bibitem{CascadedOMIT2015Fan}L. Fan, K. Y. Fong, M. Poot, and H. X. Tang, "Cascaded optical transparency in multimode-cavity optomechanical systems," Nat. Commun. {\bfseries 6}, 5850 (2015).
\bibitem{OMIT2014Dongchunhua}C. Dong, J. Zhang, V. Fiore, and H. Wang, "Optomechanically induced transparency and self-induced oscillations with Bogoliubov mechanical modes," Optica {\bfseries 1}(6), 425--428 (2014).
\bibitem{CompensationKerr2016Shen}Z. Shen, C.-H. Dong, Y. Chen, Y.-F. Xiao, F.-W. Sun, and G.-C. Guo, "Compensation of the Kerr effect for transient optomechanically induced transparency in a silica microsphere," Opt. Lett. {\bfseries 41}(6), 1249--1252 (2016).
\bibitem{EIT1991Boller}K. J. Boller, A. Imamoglu, and S. E. Harris, "Observation of electromagnetically induced transparency," Phys. Rev. Lett. {\bfseries 66}(20), 2593 (1991).
\bibitem{EITScully}M. O. Scully and M. S. Zubairy, "Quantum optics," Cambridge University Press, Cambridge, UK (1997).
\bibitem{NonlinearOMIT2013Marquardt}A. Kronwald, and F. Marquardt, "Optomechanically induced transparency in the nonlinear quantum regime," Phys. Rev. Lett. {\bfseries 111}(13), 133601 (2013).
\bibitem{HigherOrderOMIT2012Xiong}H. Xiong, L.-G. Si, A.-S. Zheng, X. Yang, and Y. Wu, "Higher-order sidebands in optomechanically induced transparency," Phys. Rev. A {\bfseries 86}(1), 013815 (2012).
\bibitem{HigherOrderOMIT2016Jiao}Y. Jiao, H. L\"{u}, J. Qian, Y. Li, and H. Jing, "Nonlinear optomechanics with gain and loss: amplifying higher-order sideband and group delay," New J. Phys. {\bfseries 18}, 083034 (2016).
\bibitem{HigherOrderKerr2018Jiao}Y. F. Jiao, T. X. Lu, and H. Jing, "Optomechanical second-order sidebands and group delays in a Kerr resonator," Phys. Rev. A {\bfseries 97}(1), 013843 (2018).
\bibitem{TwoColorOMIT2014Wang}H. Wang, X. Gu, Y. Liu, A. Miranowicz, and F. Nori, "Optomechanical analog of two-color electromagnetically induced transparency: Photon transmission through an optomechanical device with a two-level system," Phys. Rev. A {\bfseries 90}(2), 023817 (2014).
\bibitem{MultiOMIT2018Ullah}K. Ullah, H. Jing, and F. Saif, "Multiple electromechanically-induced-transparency windows and Fano resonances in hybrid nano-electro-optomechanics," Phys. Rev. A {\bfseries 97}(3), 033812 (2018).
\bibitem{SpiningOMIT2017Lv}H. L\"{u}, Y. Jiang, Y. Z. Wang, and H. Jing, "Optomechanically induced transparency in a spinning resonator," Photonics Res. {\bfseries 5}(4), 367--371 (2017).
\bibitem{PTOMIT2015Jing}H. Jing, \c{S}. K. \"{O}zdemir, Z. Geng, J. Zhang, X. Y. L\"{u}, B. Peng, L. Yang, and F. Nori, "Optomechanically-induced transparency in parity-time-symmetric microresonators," Sci. Rep. {\bfseries 5}, 9663 (2015).
\bibitem{Lightstorage2013Fiore}V. Fiore, C. Dong, M. C. Kuzyk, and H. Wang, "Optomechanical light storage in a silica microresonator," Phys. Rev. A {\bfseries 87}(2), 023812 (2013).
\bibitem{OMITCooling2014Guo}Y. Guo, K. Li, W. Nie, and Y. Li, "Electromagnetically-induced-transparency-like ground-state cooling in a double-cavity optomechanical system," Phys. Rev. A {\bfseries 90}(5), 053841 (2014).
\bibitem{OMITCooling2014Ojanen}T. Ojanen and K. B{\o}rkje, "Ground-state cooling of mechanical motion in the unresolved sideband regime by use of optomechanically induced transparency," Phys. Rev. A {\bfseries 90}(1), 013824 (2014).
\bibitem{OMITCooling2015Liu}Y.-C. Liu, Y.-F. Xiao, X.-S. Luan, and W. C. Wei, "Optomechanically-induced-transparency cooling of massive mechanical resonators to the quantum ground state," Sci. China-Phys. Mech. Astron. {\bfseries 58}(5), 050305 (2015).
\bibitem{PrecisionMeasurement2012Zhang}J. Q. Zhang, Y. Li, M. Feng, and Y. Xu, "Precision measurement of electrical charge with optomechanically induced transparency," Phys. Rev. A {\bfseries 86}(5), 053806 (2012).
\bibitem{PrecisionMeasurement2017Xiong}H. Xiong, L.-G. Si, and Y. Wu, "Precision measurement of electrical charges in an optomechanical system beyond linearized dynamics," Appl. Phys. Lett. {\bfseries 110}(17), 171102 (2017).
\bibitem{CoupledCavity2012Fan}J. Fan, and L. Zhu, "Enhanced optomechanical interaction in coupled microresonators," Opt. Express {\bfseries 20}(18), 20790 (2012).
\bibitem{UCPB2013Xu}X.-W. Xu and Y.-J. Li, "Antibunching photons in a cavity coupled to an optomechanical system," J. Phys. B: At. Mol. Opt. Phys. {\bfseries 46}(3), 035502 (2013).
\bibitem{OMITE2016Li}G. Li, X. Jiang, S. Hua, Y. Qin, and M. Xiao, "Optomechanically tuned electromagnetically induced transparency-like effect in coupled optical microcavities," Appl. Phys. Lett. {\bfseries 109}(26), 261106 (2016).
\bibitem{PTdevice2014Peng}B. Peng, \c{S}. K. \"{O}zdemir, F. Lei, F. Monifi, M. Gianfreda, G. L. Long, S. Fan, F. Nori, C. M. Bender, and L. Yang, "Parity-time-symmetric whispering-gallery microcavities," Nat. Phys. {\bfseries 10}(5), 394--398 (2014).
\bibitem{PTdevice2014Chang}L. Chang, X. Jiang, S. Hua, C. Yang, J. Wen, L. Jiang, G. Li, G. Wang, and M. Xiao, "Parity-time symmetry and variable optical isolation in active-passive-coupled microresonators," Nat. Photonics {\bfseries 8}(7), 524--529 (2014).

\bibitem{PTChaos2015LvXin}X.-Y. L\"{u}, H. Jing, J.-Y. Ma, and Y. Wu, "$\mathscr{PT}$-symmetry-breaking chaos in optomechanics," Phys. Rev. Lett. {\bfseries 114}(25), 253601 (2015).
\bibitem{PTCOM2016Xu}H. Xu, D. Mason, L. Jiang, and J. G. E. Harris, "Topological energy transfer in an optomechanical system with exceptional points," Nature {\bfseries 537}(7618), 80--83 (2016).
\bibitem{HighOrderEPPhononCooling2017Jing}H. Jing, \c{S}. K. \"{O}zdemir, H. L\"{u}, and F. Nori, "High-order exceptional points in optomechanics," Sci. Rep. {\bfseries 7}, 3386 (2017).
\bibitem{PT1998Bender}C. M. Bender and S. Boettcher, "Real spectra in non-Hermitian Hamiltonians having $\mathscr{P}\mathscr{T}$ symmetry," Phys. Rev. Lett. {\bfseries 80}(24), 5243--5246 (1998).
\bibitem{PT2007Bender}C. M. Bender, "Making sense of non-Hermitian Hamiltonians," Rep. Prog. Phys. {\bfseries 70}(6), 947--1018 (2007).
\bibitem{PT2013Bender}C. M. Bender, M. Gianfreda, \c{S}. K. \"{O}zdemir, B. Peng, and L. Yang, "Twofold transition in $\mathscr{PT}$-symmetric coupled oscillators," Phys. Rev. A {\bfseries 88}(6), 062111 (2013).
\bibitem{PTReview2016Konotop}V. V. Konotop, J. Yang, and D. A. Zezyulin, "Nonlinear waves in $\mathscr{PT}$-symmetric systems," Rev. Mod. Phys. {\bfseries 88}(3), 035002 (2016).
\bibitem{PTReview2018ELGanainy}R. El-Ganainy, K. G. Makris, M. Khajavikhan, Z. H. Musslimani, S. Rotter, and D. N. Christodoulides, "Non-Hermitian physics and $\mathscr{PT}$ symmetry," Nat. Phys. {\bfseries 14}(1), 11--19 (2018).
\bibitem{PTReview2017Feng}L. Feng, R. El-Ganainy, and L. Ge, "Non-Hermitian photonics based on parity-time symmetry," Nat. Photonics {\bfseries 11}(12), 752--762 (2017).
\bibitem{LITandEP2014Peng}B. Peng, \c{S}. K. \"{O}zdemir, S. Rotter, H. Yilmaz, M. Liertzer, F. Monifi, C. M. Bender, F. Nori, and L. Yang, "Loss-induced suppression and revival of lasing," Science {\bfseries 346}(6207), 328--332 (2014).
\bibitem{LIT2009Guo}A. Guo, G. J. Salamo, D. Duchesne, R. Morandotti, M. Volatier-Ravat, V. Aimez, G. A. Siviloglou, and D. N. Christodoulides, "Observation of $\mathscr{PT}$-symmetry breaking in complex optical potentials," Phys. Rev. Lett. {\bfseries 103}(9), 093902 (2009).
\bibitem{Tip2010Zhu}J. Zhu, \c{S}. K. \"{O}zdemir, L. He, and L. Yang, "Controlled manipulation of mode splitting in an optical microcavity by two Rayleigh scatterers," Opt. Express {\bfseries 18}(23), 23535--23543 (2010).
\bibitem{IOR1985Gardiner}C. W. Gardiner and M. J. Collett, "Input and output in damped quantum systems: Quantum stochastic differential equations and the master equation," Phys. Rev. A {\bfseries 31}(6), 3761--3774 (1985).
\bibitem{ReversedEIT}T. Oishi and M. Tomita, "Inverted coupled-resonator-induced transparency," Phys. Rev. A {\bfseries 88}(1), 013813 (2013).
\bibitem{PTReview2014Zyablovsky}A. A. Zyablovsky, A. P. Vinogradov, A. A. Pukhov, A. V. Dorofeenko, and A. A. Lisyansky, "$\mathscr{PT}$-symmetry in optics," Phys. Usp. {\bfseries 57}(11), 1063--1082 (2014).
\bibitem{EPSensing2017Hodaei}H. Hodaei, A. U. Hassan, S. Wittek, H. Garcia-Gracia, R. El-Ganainy, D. N. Christodoulides, and M. Khajavikhan, "Enhanced sensitivity at higher-order exceptional points," Nature {\bfseries 548}(7666), 187--191 (2017).
\bibitem{EPSensing2017Chen}W. Chen, \c{S}. K. \"{O}zdemir, G. Zhao, J. Wiersig, and L. Yang, "Exceptional points enhance sensing in an optical microcavity," Nature {\bfseries 548}(7666), 192--196 (2017).
\bibitem{EPSensor2018Hassan}A. U. Hassan, H. Hodaei, D. N. Christodoulides and M. Khajavikhan, "Exceptional points: an emerging tool for sensor applications," Optics \& Photonics News, {\bfseries 2018}(01), 20--22 (2018).
\bibitem{NLEP2012Konotop}D. A. Zezyulin and V. V. Konotop, "Nonlinear modes in finite-dimensional $\mathscr{PT}$-symmetric systems," Phys. Rev. Lett. {\bfseries 108}(21), 213906 (2012).
\end{thebibliography}
\end{document}